\documentclass[a4paper,cites]{article}
\usepackage{graphicx}
\usepackage{amsmath}
\usepackage{graphicx,color,psfrag}

\title{Strain bursts in plastically deforming Molybdenum micro- and nanopillars}

\begin{document}

\def \d{{\rm d}}

\maketitle

\begin{center}
{\sc M. Zaiser$^1$, J. Schwerdtfeger$^1$, A.S. Schneider$^2$,\\ 
C.P. Frick $^2$, B.G. Clark$^2$ and P.A. Gruber$^{2,3}$, and E. Arzt$^{2,4}$}\\[.5cm]

$^1$ The University of Edinburgh, Institute for Materials and
Processes,\\
The King's Buildings, Sanderson Building, Edinburgh EH9 3JL, UK\\
$^2$ Max-Planck-Institut f\"ur Metallforschung, Heisenbergstrasse 3,\\
70569 Stuttgart, Germany \\
$^3$ Universit\"at Karlsruhe, Institut f\"ur Zuverl\"assigkeit \\
von Bauteilen und Systemen, Kaiserstr. 12, 76131 Karlsruhe, Germany\\[.5cm]
$^4$ Leibniz Institute for New Materials, Campus Building D2 2, 66123 Saarbr\"ucken, Germany \\[.5cm]
\end{center}

\begin{abstract}
Plastic deformation of micron and sub-micron scale specimens is characterized by intermittent sequences
of large strain bursts (dislocation avalanches) which are separated by regions of near-elastic 
loading. In the present investigation we perform a statistical characterization of strain bursts 
observed in stress-controlled compressive deformation of monocrystalline Molybdenum micropillars. We characterize 
the bursts in terms of the associated elongation increments and peak deformation rates, and demonstrate that these 
quantities follow power-law distributions that do not depend on specimen orientation or stress rate. We also 
investigate the statistics of stress increments in between the bursts, which are found to be Weibull distributed
and exhibit a characteristic size effect. We discuss our findings in view of observations of deformation bursts
in other materials, such as face-centered cubic and hexagonal metals. 
\end{abstract}

\section{Introduction}

On microscopic and mesoscopic scales, plastic deformation of crystalline solids proceeds as an intermittent series of strain bursts ('slip avalanches'). Indirect evidence of such bursts has been provided by systematic acoustic emission (AE) studies of Weiss and co-workers on ice \cite{WEI01,RIC05}, hcp metals \cite{RIC06}, and fcc metals \cite{WEI08}. These studies indicate that the AE signals of plastically deforming crystals consist of discrete bursts separated by quiescent intervals of low AE activity. The energies $E$ (amplitude square integrated over the duration of a burst) and peak amplitudes $A$ of the AE bursts exhibit a huge scatter; their statistics is characterized by scale-free (power law) distributions, with probability 
density functions $p(E) \propto E^{-\kappa_E}$ and $p(A) \propto A^{- \kappa_A}$ that are well described as power laws with material-independent exponents $\kappa_E \approx 1.5$ and $\kappa_A \approx 2$, extending over up to 8 decades with no apparent cut-off.

Dimiduk and co-workers confirmed temporal intermittency of plastic flow by direct observation of strain bursts during compressive deformation of micropillars machined out of Ni single crystals \cite{DIM06}. In these experiments, the elongation vs. time curves observed under stress-controlled loading were characterized by an intermittent sequence of deformation jumps, with elongation increments $\Delta l$ that exhibited a scale-free distribution $p(\Delta l) \propto \Delta l^{-\kappa_l}$
where $\kappa_l \approx 1.5$. Recently, Ngan observed bursts with similar statistical characteristics in creep deformation of aluminum micropillars under constant stress conditions \cite{NGA07}. These findings can be directly related to the acoustic emission results if one assumes that a fixed fraction of the work done by the external forces during an elongation jump is released in the form of acoustic energy. 

Theoretically, the formation of intermittent deformation bursts has been modelled using two-dimensional \cite{MIG01} and three-dimensional \cite{CSI07} discrete dislocation dynamics simulations, as well as various types of continuum models \cite{ZAI03,KOS04,ZAI05,ZAI06}. In the discrete simulations, the stochastic nature of the deformation process is directly 'inherited' from the statistical choice of the initial dislocation configuration, which in turn reflects the variability of the initial microstructure of real specimens. In the continuum models, statistical heterogeneity needs to be explicitly incorporated into the constitutive equations, e.g. in terms of fluctuations in the local flow stresses or energy dissipation rates. While less 'realistic' than discrete dislocation simulations, such models provide a conceptual framework for understanding the origin of the scale-free avalanche dynamics which can be related to a depinning-like transition between an elastic and a plastically deforming phase (`yielding transition'). A comprehensive overview of experimental and theoretical results has been given by Zaiser \cite{ZAI06b}.

In spite of all these investigations, open questions remain. The relations between different characteristics of strain bursts, such as the associated strain or elongation increments, the burst durations and the peak strain rates, have remained largely unexplored. Other open issues concern the statistics of stress increments between bursts and the correlation between stress increments and burst sizes. Furthermore, practically all the experimental evidence has been gathered on materials (fcc and hcp metals, ice close to its melting point) where the motion of dislocations is governed by their mutual interactions. Molybdenum (Mo), on the other hand, is a bcc metal where dislocation interactions with the crystal lattice (Peierls stresses) may have a crucial influence on the deformation behavior: Below the so-called knee temperature ($\sim$ 500-550 K for Mo \cite{SES78}), the plastic deformation of bulk bcc metals is controlled by the nucleation and motion of kinks on screw dislocations. These materials exhibit therefore a strong temperature and strain-rate dependence of the flow stress (for a detailed overview of the deformation behavior of bcc metals, see \cite{SES78,SEE02}). Since the motion of dislocations is at low temperatures governed by their interactions with the crystal lattice, it has been argued that collective behavior, and hence strain bursts, may be suppressed in this temperature regime \cite{WEI08}. 

In the present paper, we adress these open questions by investigating strain bursts observed in Molybdenum (Mo) micropillars that are deformed at room temperature in compression under load control. In the following sections we first describe the experimental procedure and the methods used for characterizing the strain bursts. We then discuss the results of our analysis in view of the statistics of burst sizes and stress increments, and the relations between burst strain and peak strain rate. We conclude with a comparison of our findings with theoretical and experimental results on strain bursts in other materials.

\section{Experimental}

\subsection{Specimen preparation and mechanical testing}

Zone refined Mo single crystals were oriented using Laue diffraction, and disk-shaped samples of approximately 3 mm height and 10 mm diameter with disk normals pointing along [100] and [235] lattice directions were cut by spark erosion. The disk surfaces were then mechanically polished using 6, 3 and 1 $\mu$m diamond suspensions and subsequently electro-polished for 60 seconds using a mixture of 610 ml Methanol and 85 ml H$_2$SO$_4$ at a current of 1-2 Amperes. The samples were mounted on a custom machined aluminum holder for testing, and their orientation was confirmed by electron 
backscatter diffraction.  

The specific process used in fabricating micropillars from the oriented samples is very similar to the method of Frick et al. \cite{FRI07}. Free-standing pillars of tapered shape were fabricated using a dual focused ion beam (FIB) and scanning electron microscope (SEM) (FEI Nova 600 NanoLab DualBeam$^{\rm TM}$). The as-machined pillars were deformed in compression at ambient pressure and temperature by using a MTS XP nanoindenter system equipped with a sapphire conical indenter with a flat 10 $\mu$m diameter tip. The loading rates varied between 4 and 60 $\mu$N/sec, depending on pillar diameter. Geometrical parameters (top diameter $d_{\rm t}$, bottom diameter $d_{\rm b}$, and length $l$ of the analysed pillars) and loading rates  are compiled in Table 1.

\begin{figure}
\begin{center}
\includegraphics{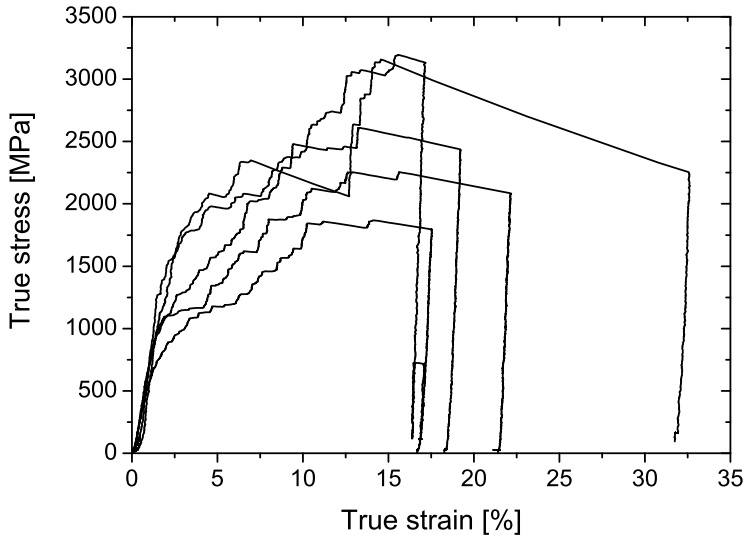}
\end{center}
\caption{Stress-strain curves of [100] oriented Mo micropillars \cite{SCH08} 
(Specimens 2$_{\rm nr}$, 5$_{\rm nr}$, 6$_{\rm nr}$, 7$_{\rm nr}$, 9$_{\rm nr}$, and 10$_{\rm nr}$ in Table 1).}
\label{stresstrain}
\end{figure}

Deformation experiments were performed at a control rate of 500 Hz with a data storage rate of 25 Hz, i.e., data were recorded at intervals $\Delta t = 0.04$ s.  Tests were typically performed with two intermediate unloading and reloading cycles at about 2.5\% and 5\% strain in order to observe the linear elastic response and transient loading/unloading behavior of the pillars. The intermediate unloading may influence the strain burst statistics: Most of the deformation occurs during the largest bursts (see Figure \ref{stresstrain}) and it is thus very likely that these are truncated by unloading (in fact, sometimes the burst continued during unloading or burstlike deformation resumed at a reduced stress level upon reloading). Therefore, a set of [100] oriented pillars (labelled with the subscript 'nr' in Table 1) were deformed without intermediate unloading. Analysing these separately allows us to assess the influence of unloading on the burst statistics. 

\begin{figure}
\begin{center}
\includegraphics{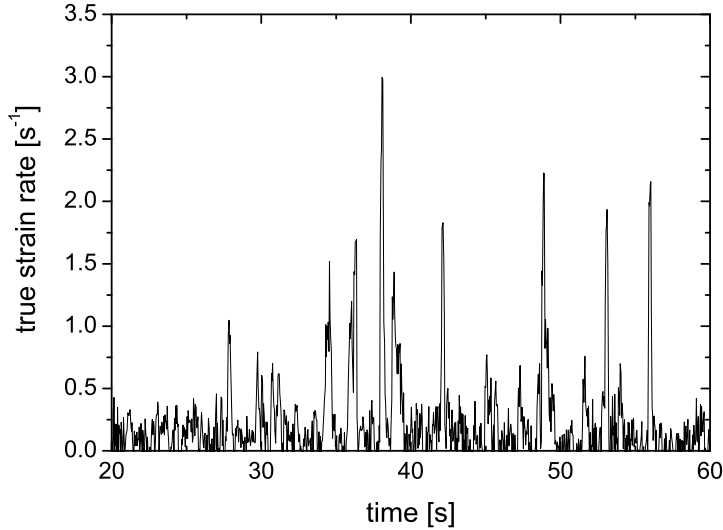}
\end{center}
\caption{Strain rate vs. time signal during deformation of a [100] oriented Mo 
micropillar (Specimen 10$_{\rm nr}$, lowermost curve in Figure 1).}
\label{strainrate}
\end{figure}

The flow stresses of the investigated samples increase with decreasing sample size. A study of this size effect has been published elsewhere \cite{SCH08}. Here, we focus exclusively on the intermittent nature of the deformation process.  
As can be seen from Figure \ref{stresstrain}, the deformation curves are characterized by an irregular sequence of large strain 
bursts visible as steps on the stress vs. strain or elongation vs. time curves. During the bursts, which typically lasted less than a second, deformation rates were high (peak strain rates $>$ 1 s$^{-1}$). The elongation rate signals have the typical signature of a 'crackling noise' (Figure \ref{strainrate}) \cite{SET01}, i.e., they are composed of discrete bursts of widely varying magnitude.   

\subsection{Data analysis}

Strain bursts were characterized in terms of their size (defined as the elongation increment between the beginning and the end of the burst), duration, and peak elongation rate. To define a burst, the elongation vs. time signals $l(t)$ were first conditioned by performing a running average over an averaging time interval $\Delta t_{\rm av}$. This served to eliminate high-frequency noise resulting from the deformation setup. The averaged signals $\bar{l}(t)$ were then differentiated using a simple central difference scheme, and the resulting elongation rate signals $\d_t\bar{l}(t)$ were broken into bursts by thresholding: A strain burst was associated with a time interval $[t^i_1,t^i_2]$ such that $\d_t\bar{l}(t) > \dot{l}_{\rm thresh}$ for all $t \in [t^i_1,t^i_2]$ and $\d_t\bar{l}(t) < \dot{l}_{\rm thresh}$ for $t = t^i_1 - \Delta t$ and for $t = t^i_2 + \Delta t$. The burst duration was then defined as $T^i := t^i_1-t^i_2$, the burst elongation as $S^i = l(t^i_2) - l(t^i_1)$, the time-averaged peak elongation rate as $\dot{L}^i_{\rm pav} := {\rm max}\, [d_t\bar{l}(t)]$
for $t \in [t^i_1,t^i_2]$, and the true peak elongation rate as $\dot{L}^i_{\rm p} := {\rm max}\, [d_t l(t)]$
for $t \in [t^i_1,t^i_2]$. The burst initiation stress was defined as $\sigma^i := \sigma(t^i_1)$, and the stress increment as $\Delta \sigma^i := \sigma^{i}-\sigma^{i-1}$. On some rare occasions, bursts occurred during intermediate unloading or reloading, leading to negative stress increments. These bursts were discarded from the stress increment statistics. 

In our analysis we used the standard parameters $\Delta t_{\rm av} = 0.8$ s and $\dot{l}_{\rm thresh} = 0.5 $ nm/s. For these parameters, a typical specimen of $0.5\, \mu$m diameter yielded between 50 and 100 bursts, most of them small. Since this is not sufficient for a meaningful statistical analysis, we grouped specimens of the same orientation into size classes with typically 6-8 specimens in each class. Those [100] specimens that were deformed without intermediate stress relaxation are grouped separately such that the influence of intermediate unloading on the strain burst statistics can be assessed. The class partition is shown in Table 1. For the bursts obtained from all specimens in a given class, probability distributions $p(S)$ were determined by logarithmically binning the $S^i$ data. This is appropriate for power-law distributed data where logarithmic binning may significantly improve the statistics in the regime of large events without introducing spurious cut-off effects.  
Stress increments, on the other hand, were found to be Weibull distributed. In this case, since the data scatter around a characteristic value, logarithmic binning makes little sense. Instead, we base our statistical analysis on the cumulative distribution $P_{>}(\Delta \sigma)$ as determined from the ordered sequence of the $\Delta \sigma^i$: $P_{>}(\Delta\sigma_n) \approx n/(N+1)$ where $N$ is the total number of stress increments and $\Delta\sigma_n$ the $n$th member in the descending sequence. 

To ensure that the burst statistics do not significantly depend on the signal conditioning and thresholding parameters $\Delta t_{\rm av}$ and $\dot{l}_{\rm thresh}$, we performed a systematic parameter study by varying these parameters in the ranges $0.2$ s $ \le \Delta t_{\rm av} \le 2.4$ s and $0.25$ nm/s $\le \dot{l}_{\rm thresh} \le 2.5 $ nm/s, and studying the corresponding changes in the $p(S)$ probability distribution for 'small' [100] oriented pillars (class [100]S, Table 1).

\section{Results and Discussion}

\begin{figure}[tb]
\begin{center}
\includegraphics{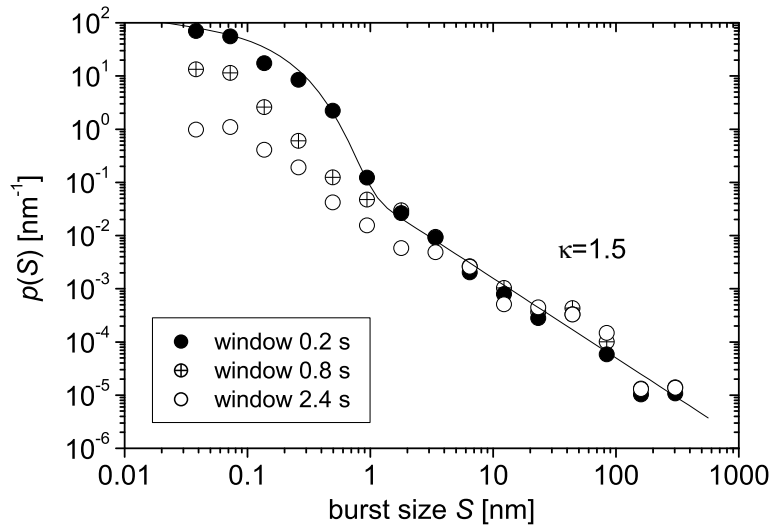}
\end{center}
\caption{Strain burst size distributions for 'small' [100] 
oriented Mo micropillars (class [100]S in Table 1), determined with an elongation rate
threshold of 0.02 nm/s and different sizes of the averaging window; full line:
fit function  $p(x) = 100 \exp[-x/0.12] +0.05 x^{-1.5}$}
\label{windows}
\end{figure}

We first investigate to which extent the statistics of strain burst sizes is influenced by the parameters used for smoothing and thresholding the raw elongation rate signals. Figure \ref{windows} shows distributions obtained for 'small' [100] crystals using three different sizes of the averaging window. 
For short averaging windows, the distribution of burst sizes exhibits two distinct regimes: At small burst sizes, the burst size distribution has a 'hump' which decays exponentially, whereas at large sizes, the exponential decay is replaced by a power-law tail. We may associate these two regimes with two different physical processes, {\em viz} on the one hand the high-frequency noise of the deformation setup which produces a large number of small 'bursts' -- in fact, just irregular oscillations of the deformation machine -- and on the other hand the collective dynamics of the dislocation system which produces intermittent large bursts of plastic deformation activity with a power-law size distribution. What is important is that the high-frequency noise of the machine does not mask the power-law scaling since the amplitude of the machine-induced elongation fluctuations is limited to values less than 1 nm. By increasing the length of the averaging window, we can suppress this exponential 'hump' while the power-law part of the distribution remains unchanged -- in fact, the length of the scaling regime increases and reaches a maximum at a window length of 0.8 s which we choose as our default value. If the original signal is averaged over even larger times, the size distribution of large bursts remains unchanged but the length of the scaling regime decreases again since smaller bursts are 'washed out' as their peak elongation rates fall below the threshold. 

\begin{figure}[bt]
\begin{center}
\includegraphics{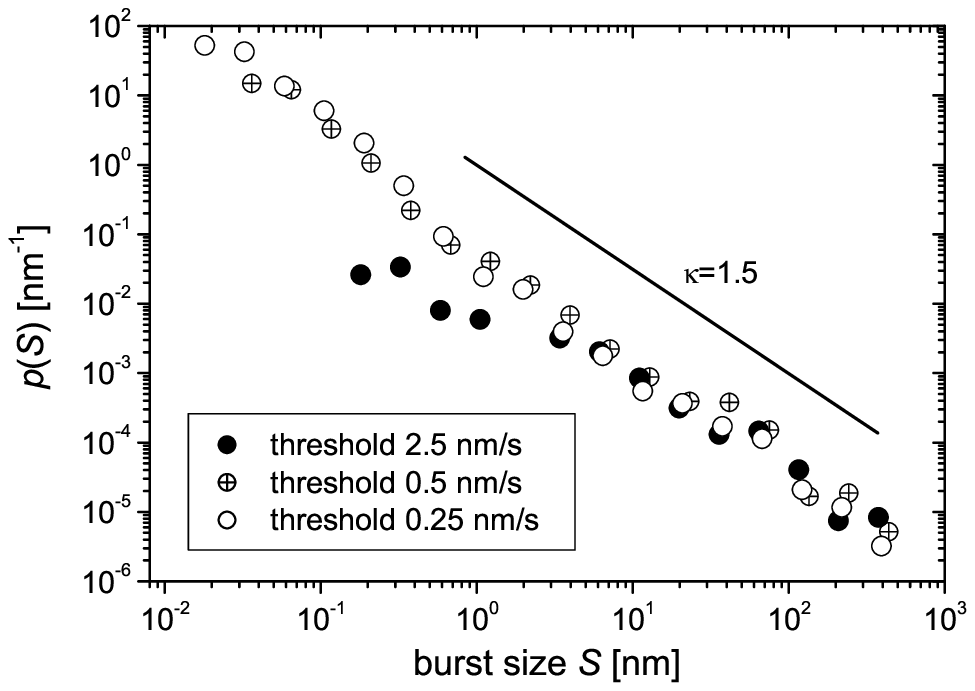}
\end{center}
\caption{Strain burst size distributions for 'small' [100] 
oriented Mo micropillars (class [100]S in Table 1), determined with  
different threshold values of the elongation rate and a window size
of 0.8 s.}
\label{thresholds}
\end{figure}

Figure \ref{thresholds} shows the dependence of the burst size distribution on the imposed elongation rate threshold. For small thresholds, the distribution is practically independent on threshold, while a threshold  substantially above our 
default value of 0.5 nm/s eliminates small bursts but leaves the size distribution of large bursts practically unchanged. Crucially, neither the size of the averaging window nor the choice of the threshold seem to have any appreciable influence on the power-law scaling of the burst size distribution in the large-burst regime (elongation increments larger than approximately 1 nm). This robustness of the procedure indicates that it is indeed viable to envisage our elongation rate signals as 'crackling noise' composed of discrete events. 

\begin{figure}
\begin{center}
\includegraphics{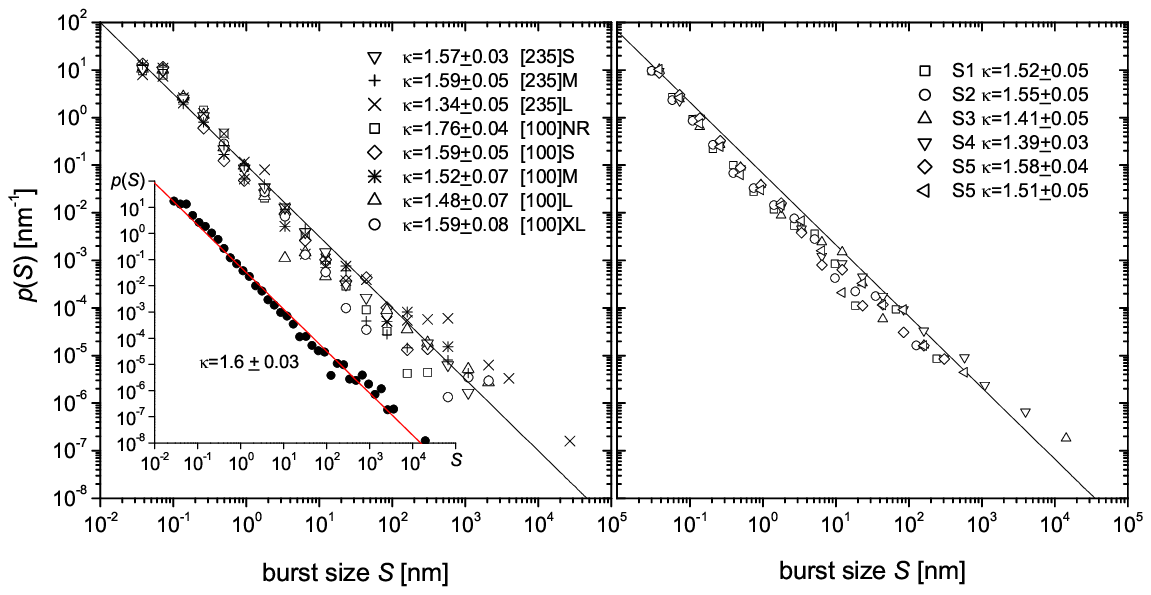}
\end{center}
\caption{Left: Compilation of strain burst size distributions for the different specimen classes, the legend shows
the $\kappa$ values for each class; inset: aggregated distribution of burst sizes from all classes; right: surrogate data
(distributions determined from sets of 400 random numbers $S$ drawn from a distribution $p(S) \propto S^{-1.5}$ with $S_{\rm min} = 0.02$); full lines: $\kappa = 1.5$.}
\label{psall}
\end{figure}

Probability distributions of strain burst sizes for the different specimen classes are shown in Figure \ref{psall} (left). All distributions can with reasonable accuracy be described as power laws:
\begin{equation}
p(S) \propto S^{-\kappa}\;.
\end{equation}
Least-square fits to the logarithmically binned data yield values of $1.34 \le \kappa \le 1.76$, and compiling bursts from all specimens and determining the overall size distribution as shown in the inset of Figure \ref{psall} (left) yields $\kappa = 1.6 \pm 0.03$. No systematic dependency of the exponent $\kappa$ on pillar orientation or pillar size can be detected. Even though intermediate unloading is expected to truncate some of the largest bursts, [100] oriented specimens deformed without unloading  do not exhibit larger bursts than those from the other groups - if anything, the above average exponent $\kappa \approx 1.76$ for the [100]NR class suggests the opposite. Differences between the distributions obtained for different specimen classes should not be over-interpreted -- they may simply reflect statistical scatter inherent in the not very large size of the datasets which comprise typically some 400 bursts for each specimen class. To illustrate this point, we show on the right-hand side of Figure \ref{psall} simulated $p(S)$ distributions determined from 6 sets of surrogate data, each consisting of 400 random numbers drawn independently from a distribution $p(S) \propto S^{-1.5}$. As can be seen, the scatter of the exponents determined from these sets, the scatter in the data ranges, and the error of the linear least-square fits are all comparable with the corresponding values for the experimental datasets. This illustrates the intrinsic problems encountered in determining distribution parameters from limited sets of data. 

\begin{figure}
\begin{center}
\includegraphics{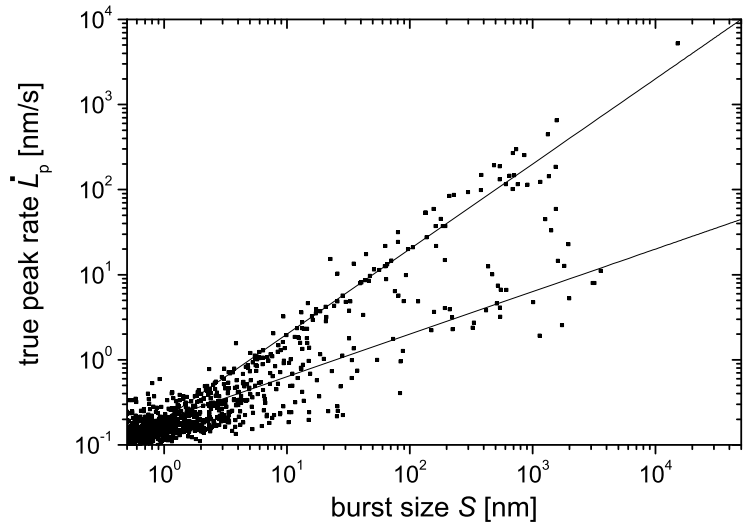}
\end{center}
\caption{Relation between burst size $S$ and true peak elongation rate $\dot{L}_{\rm p}$; upper straight line: $\dot{L}_{\rm p}\propto S$, lower straight line: $\dot{L}_{\rm p}\propto S^{1/2}$.}
\label{sizerate}
\end{figure}

We now proceed to investigate other burst characteristics, {\em viz} the burst durations and peak elongation rates. 
Unfortunately, the intrinsic burst durations may be well below the size $\Delta t_{\rm av}$ of our averaging window. As a consequence, all large bursts determined from the averaged elongation rate signal have approximately the same duration which is roughly proportional to $\Delta t_{\rm av}$. Hence, the burst durations as determined from the averaged signals are no longer good characterizers of the bursts and, for evident reasons, the same is true for the peak rates of the time-averaged signals which decrease with increasing $\Delta t_{\rm av}$. The true peak elongation rate $\dot{L}_{\rm p}$, on the other hand, represents an intrinsic property of the bursts that is not affected by time averaging. There is a strong statistical correlation between $\dot{L}_{\rm p}$ and burst size $S$ (correlation coefficient $r > 0.9$) but it it is not easy to establish a clear-cut mathematical relation between the two quantities. This is seen from Figure \ref{sizerate} which shows $\dot{L}_{\rm p}$ vs. $S$ values for all large bursts. While part of the observed bursts seem to exhibit peak rates that are approximately proportional to the burst sizes ($\dot{L}_{\rm p}\propto s$, upper straight line in Figure \ref{sizerate}), other data seem to suggest a proportionality to the square root of the burst sizes $\dot{L}_{\rm p}\propto S^{1/2}$, lower straight line in Figure \ref{sizerate}). Fitting a power law to all the data yields $\dot{L}_{\rm p} \propto S^{0.8}$ which badly represents either group. Interpretation of these findings is further complicated by the fact that the two behaviors do not represent different specimen classes -- rather, bursts from one and the same specimen may be found both near the upper and the lower straight lines.

To assess the degree of correlation between the burst size and the magnitude of the preceding and following stress increments, the respective correlation coefficients were evaluated separately for the 6 datasets in class [100]S, and the mean correlation coefficient as well as the variance of the $r$ values were determined. The results ($r = 0.03 \pm 0.24$ for the correlation coefficient between burst size and magnitude of the preceding stress increment, and $r = 0.11 \pm 0.28$ for the correlation  between burst size and magnitude of the following stress increment) do not indicate any statistically significant correlation. There is also no statistically significant correlation between the sizes of successive bursts ($r = 0.02 \pm 0.05$). 

\begin{figure}
\begin{center}
\includegraphics{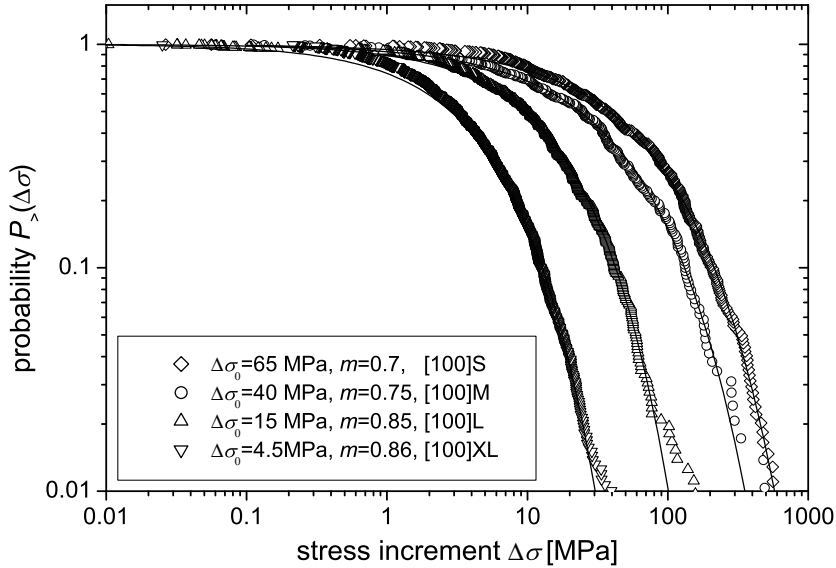}
\end{center}
\caption{Distributions of stress increments for specimen classes [100]S, [100]M, [100]L, and [100]XL; full lines: Weibull fits, for parameters see inset in figure.}
\label{stressinc}
\end{figure}

The statistics of stress increments differs substantially from the statistics of burst sizes: Instead of scale-free power laws we find distributions with a characteristic scale that depends on specimen size. Figure \ref{stressinc} shows cumulative distributions $P_>(\Delta \sigma)$ (probability to find a stress increment larger than $\Delta \sigma$) for the specimen classes [100]S, [100]M, [100]L, and [100]XL. The data can be well fitted by Weibull distributions, 
\begin{equation}
P_>(\Delta \sigma) = \exp\left[-\left(\frac{\Delta \sigma}{\Delta \sigma_0}\right)^m\right]\;,
\end{equation}
where $m$ is the Weibull modulus and the stress parameter $\Delta \sigma_0$ defines the characteristic stress increment. Parameters for the distributions are shown in the legend; the Weibull modulus $m$ which determines the 
width of the distribution is approximately the same for all distributions, but the stress parameter decreases with increasing specimen size. 

\begin{figure}
\begin{center}
\includegraphics{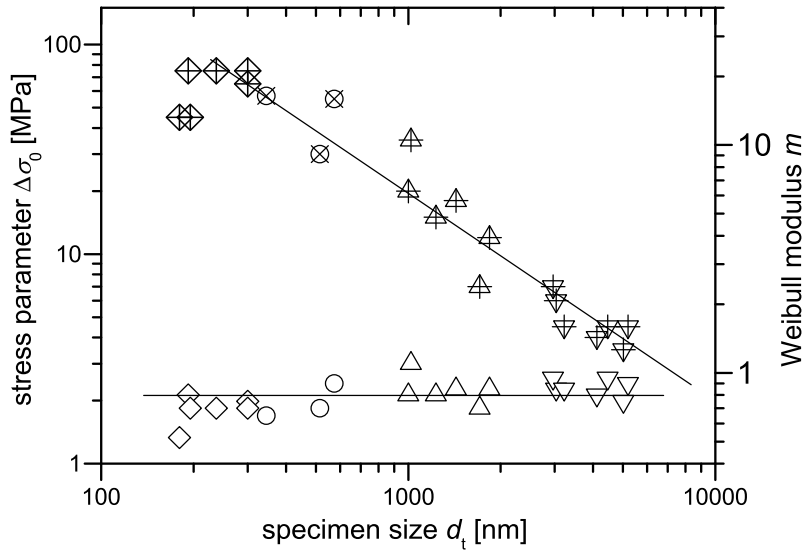}
\end{center}
\caption{Parameters of Weibull fits to stress increment distributions determined for individual specimens; different
symbol shapes distinguish different specimen classes ({\huge$\diamond$} [100]S, $\bigcirc$ [100]M, $\bigtriangleup$ [100]L, $\bigtriangledown$ [100]XL); open symbols: Weibull moduli; cross-center symbols: stress parameters.}
\label{weibull}
\end{figure}

This is further illustrated in Figure \ref{weibull} where parameters $m$ and $\Delta \sigma_0$ of stress increment distributions obtained from individual specimens are plotted against specimen size. It is clearly seen that the Weibull moduli $m \approx 0.75$ do not depend significantly on specimen size, whereas the stress parameters $\Delta \sigma_0$ (and, accordingly, the average stress increments between bursts) decrease approximately in inverse proportion with specimen diameter $d_{\rm t}$. This implies that, in larger specimens, smaller stress increments are needed to trigger strain bursts -- an obvious result since in larger specimens we expect to find a larger number of weak regions or sources that can be activated in any given stress interval. A more quantitative analysis is, however, hampered by the fact that the characteristic stress increments depend on the averaging and thresholding parameters used in our data analysis: shorter averaging times $\Delta t_{\rm av}$ and smaller threshold values $\dot{l}_{\rm thresh}$ lead to the identification of a larger number of small 'bursts' and a proportional reduction of the characteristic stress increment $\Delta \sigma_0$. Therefore, without a method to clearly distinguish between machine-induced noise and the smaller bursts that result from collective dislocation motion, it is difficult to draw quantitative conclusions from the observed size dependence of the $P(\Delta \sigma)$ distributions. 

\section{Conclusions}

Our investigation provides an example of plasticity behaving as a 'crackling noise' \cite{SET01}, with intermittent bursts of activity characterized by scale-free size distributions. For the burst sizes (elongation increments) we find a distribution $p(S) \propto S^{-1.5}$ which is in line with experimental findings on Ni micropillars \cite{DIM06} as well as 
theoretical predictions based on continuum and discrete dislocation models \cite{ZAI06}. The same theoretical models predict power-law relationships $\dot{L}_{\rm p}\propto T \propto S^{1/2}$ to hold between the peak rate, duration, and size of strain bursts. Unfortunately, owing to the need for conditioning the signal by time averaging, no useful information about the burst durations could be obtained in the present investigation, while the information regarding the relationship between burst size and peak elongation rate turned was found to be ambiguous. 

In line with previous investigations, the power-law characteristics of strain bursts seem to be little affected by specimen orientation, size, or imposed deformation rate. While theoretical investigations \cite{CSI07,ZAI07} suggest an intrinsic cut-off to the power-law scaling regime, no such cut-off could be identified in our investigation. This may be due to the fact that establishing a cut-off requires good statistics in the region of very large strain bursts, which could not be achieved in the present investigation as the total number of bursts obtained from each individual specimen was small ($<$ 100). 

The distributions of stress increments between subsequent strain bursts differ substantially from the burst size distributions. Instead of scale-free power laws, we find Weibull distributions with a characteristic stress scale (the stress parameter $\Delta \sigma_0$) that decreases approximately in inverse proportion with specimen size. However, the very presence of a characteristic scale makes the distribution parameters depend on the number of identified bursts. This dependency raises the problem of distinguishing between the effects of collective dislocation motion and the effects of machine noise, which may increase the apparent burst number by adding spurious 'bursts' of small size into the statistics. For the same reason, any conclusions based upon the observed lack of correlation between burst sizes and strain increments, or between the sizes of subsequent bursts, must be regarded with caution. 

Our investigation demonstrates for the first time the occurrence of scale-free strain bursts in a bcc metal deforming below the transition temperature, and we find that the burst characteristics are similar to those in fcc metals. This implies that even a significant Peierls stress, which is of crucial importance for the deformation properties of bulk Mo at ambient tempeature, is not sufficient to inhibit burst-like deformation. In this sense, we may conclude that the Peierls potential is irrelevant as far as the dynamics and statistics of strain bursts are concerned, and that the observed behaviour constitutes a truly universal feature of dislocation plasticity that can be observed in all kinds of crystal lattice structures. It may be mentioned that the same is not true for the size dependence of the flow stress: The size effects observed in fcc micropillars (for reference, see, e.g. \cite{DIM05}) differ substantially from those observed in the present samples \cite{SCH08}.  The universality of strain burst behavior is a key result of the present study, and it would be desirable to obtain further corroboration of this result from acoustic emission measurements on bulk bcc metals.

{\bf Acknowledgements:} We acknowledge support of the Commission
of the European Communities under contract
NEST-2005-PATH-COM-043386 and of EPSRC under Grant No. EP/E029825.

\begin{table}

\caption{Specimen orientations ([100] or [235]), geometries and loading rates.}  
\begin{tabular}{| l | l | l | l | l | l |}

\hline
Class&        No &    $d_{\rm t}$ [nm] &    $d_{\rm b}$ [nm] &     $l$ [nm] & $\dot{\sigma}$ [MPa/s] \\
\hline
\hline
[100]S  &    
                 15 &        192 &        217 &        515 &      101.1  \\
$d$=180-300 nm &
                 16 &        237 &        282 &        673 &       67.0  \\
$\langle \dot{\sigma} \rangle = 58.6$ MPa/s&        
                 17 &        300 &        363 &        691 &       84.0 \\

&                18 &        300 &        347 &        692 &       28.1 \\

&                19 &        195 &        250 &        761 &       32.7 \\

&                20 &        180 &        227 &        556 &       38.5 \\

\hline

[100]M &
        				 12 &        574 &        619 &        878 &       38.5 \\
$d$=340-600nm&
                 13 &        515 &        572 &       1030 &       19.0 \\
$\langle \dot{\sigma} \rangle = 36.9$ MPa/s &            
                 14 &        345 &        390 &        764 &       53.1 \\

\hline

[100]L  &          
				         4 &       1000 &       1160 &       2680 &       38.1 \\
$d$=1000-2000nm &
                 5 &       1430 &       1710 &       3170 &       31.1 \\
$\langle \dot{\sigma} \rangle = 29.9$ MPa/s & 
                 6 &       1230 &       1370 &       2750 &       42.0 \\

&                7 &       1020 &       1240 &       2600 &       40.7 \\

&                10 &       1840 &       2020 &       3670 &       18.8 \\

&                11 &       1710 &       1950 &       3650 &        8.7 \\

\hline
    
[100] XL &     1 &       3030 &       3700 &       7380 &        8.3 \\
$d>$ 2000nm &       2 &       2960 &       3440 &       6700 &        8.7 \\
$\langle \dot{\sigma} \rangle = 6.4$ MPa/s & 
                   3 &       3220 &       3520 &       6910 &        6.1 \\
                   
&                 21 &       5020 &       5740 &      10070 &        3.0 \\

&                 22 &       4110 &       4700 &       7220 &        4.5 \\

&                 23 &       4460 &       5050 &       9180 &        6.4 \\

&                 26 &       5100 &       5620 &       9430 &        4.9 \\

\hline

[100]NR &
         2$_{\rm nr}$ &        166 &        286 &        911 &       85.2 \\
$d$=150-435nm &    
         3$_{\rm nr}$ &        204 &        299 &        870 &       56.5 \\
$\langle \dot{\sigma} \rangle = 53.3$ MPa/s & 
         4$_{\rm nr}$ &        172 &        315 &        864 &       77.5 \\
no relaxation &
         5$_{\rm nr}$ &        192 &        295 &        775 &       62.4 \\
&
         7$_{\rm nr}$ &        393 &        550 &       1320 &       40.1 \\
&
         8$_{\rm nr}$ &        393 &        550 &       1250 &       39.8 \\
&
         9$_{\rm nr}$ &        435 &        535 &       1170 &       32.6 \\
&
        10$_{\rm nr}$ &        435 &        535 &       1190 &       32.2 \\

\hline
\hline
[235]S   &
         8 &        227 &        380 &        634 &       65.1 \\
$d$ = 200-600nm & 
         9 &        324 &        436 &        678 &       35.9 \\
$\langle \dot{\sigma} \rangle = 49.6$ MPa/s & 
        10 &        308 &        500 &       1010 &       52.5 \\

&       11 &        674 &        912 &       1580 &       44.6 \\

&       12 &        470 &        647 &       1270 &       57.2 \\

&       13 &        590 &        802 &       1320 &       36.4 \\

&       14 &        603 &        813 &       1370 &       55.7 \\

\hline
[235]M &
         1 &       1220 &       1570 &       2740 &       25.6 \\
$d$ = 650-1500nm &  
         2 &       1370 &       1690 &       2450 &       20.3 \\
$\langle \dot{\sigma} \rangle = 28.1$ MPa/s &
         3 &       1250 &       1650 &       2440 &       24.4 \\

&        4 &       1370 &       1750 &       2440 &       20.3 \\

&       15 &        694 &        860 &       1220 &       42.2 \\

&       16 &        752 &        920 &       1260 &       35.9 \\

\hline

[235]L &
         5 &       3420 &       4330 &       6380 &       19.1  \\
$d>$ 1500nm & 
         6 &       3470 &       4510 &       7060 &       18.6 \\
$\langle \dot{\sigma} \rangle = 14.2$ MPa/s &
         7 &       3460 &       4530 &       7570 &       18.6 \\

&       17 &       5660 &       7480 &      15960 &       11.0 \\

&       18 &       5820 &       7480 &      16950 &       10.4 \\

&       19 &       5810 &       7080 &      15410 &       10.4 \\

&       20 &       5580 &       7030 &      14900 &       11.3 \\

\hline

\end{tabular}

\end{table}

\end{document}